\documentstyle[aps,multicol,epsf,psfig]{revtex}
\newcommand{\ket}[1]{\left | #1 \right \rangle}

\begin{document}

\title{ Greenberger-Horne-Zeilinger paradox for continuous variables}
\author{Serge Massar and Stefano Pironio}
\address{Service de Physique Th\'eorique,
Universit\'e Libre de Bruxelles, CP 225, Bvd du Triomphe, B1050
Bruxelles, Belgium.}

\maketitle

We show how to construct states for which a
Greenberger-Horne-Zeilinger type paradox occurs if each
party measures either the position or momentum of  his particle. The
paradox can be ascribed to the anticommutation of certain
translation operators in phase space. We then rephrase the paradox in terms 
of modular and binary variables. The origin
of the paradox is then due to the fact that the associativity of addition of
modular variables is true only for c-numbers but does not hold for operators.

\begin{abstract}

\end{abstract}

%Preliminary Version -- 11 January 2001.\\
%\pacs{03.65.Bz??}

%\vspace{-1.1cm}
\begin{multicols}{2}

The Greenberger-Horne-Zeilinger (GHZ) paradox\cite{ghz,m}
exhibits in a very simple way  the
non locality inherent to quantum mechanics. The
paradox is based on the following state 
\begin{equation}
\ket{\Psi^{GHZ}} = {\ket{\uparrow_A \uparrow_B \uparrow_C} - 
\ket{\downarrow_A \downarrow_B \downarrow_C} \over \sqrt{2} }
\end{equation}
of three spin 1/2 particles shared among three spatially separated
parties A,B,C. The GHZ
state is an eigenstate of four commuting operators, the GHZ operators:
\begin{eqnarray}
\sigma_{x}^A  \sigma_{x}^B  \sigma_{x}^C \ket{\Psi^{GHZ}} &=& 
- \ket{\Psi^{GHZ}},
\nonumber\\
\sigma_{x}^A  \sigma_{y}^B  \sigma_{y}^C \ket{\Psi^{GHZ}} &=& 
+\ket{\Psi^{GHZ}},
\nonumber\\
\sigma_{y}^A  \sigma_{x}^B  \sigma_{y}^C\ket{ \Psi^{GHZ}} &=& 
+\ket{\Psi^{GHZ}},
\nonumber\\
\sigma_{y}^A \sigma_{y}^B  \sigma_{x}^C\ket{ \Psi^{GHZ}} &=& 
+\ket{\Psi^{GHZ}}
\ .
\label{spinGHZ}
\end{eqnarray}
The paradox arises when one compares the predictions of quantum
mechanics and of local hidden variable theories for local measurements of
$\sigma_x$ and $\sigma_y$ by the three parties. 
By carrying out such local measurements, one
can measure any of
the four GHZ operators 
that appear on the left hand side of 
(\ref{spinGHZ}).
Quantum mechanics
predicts that for the state $\ket{\Psi^{GHZ}}$ measuring any of
the four GHZ operators must yield a
result equal to the eigenvalue appearing on the right hand side of 
(\ref{spinGHZ}). On the other hand in a local hidden variable theory
one must, prior to the measurement, assign values to all six of the
operators $\sigma_{x,y}^j$, $j=A,B,C$. These values must be equal to
one of the eigenvalues of the operators, ie. to $\pm 1$. But it is
impossible to assign values to all six operators $\sigma_{x,y}^j$ and
be compatible with the predictions of quantum mechanics. Indeed taking
the product of the 4 GHZ equations one obtains $+1$ when the
$\sigma_{x,y}^j$ are replaced by c-numbers. Quantum mechanically one
obtains $-1$ because $\sigma_{x}^j$ and $\sigma_{y}^j$ anti-commute.

The aim of the present paper is to show how one can generalize in a
natural way the GHZ 
paradox to continuous variables such as position and momentum. 
Our analysis is inspired by the analogy
between the  EPR state for continuous variables
$\ket{\Psi^{EPR}} = \int dx \ket{x_A, -x_B}$ and the 
singlet state for discrete systems $\ket{\Psi^-} = (\ket{\uparrow_A
  \downarrow_B} - \ket{\downarrow_A
  \uparrow_B})/\sqrt{2}$. These states can be defined 
in terms of the operators of which they are eigenstates: 
$(x_A + x_B) \ket{\Psi^{EPR}} =0$ , $(p_A - p_B) \ket{\Psi^{EPR}} =0$
and
$\sigma_{z}^A \sigma_{z}^B \ket{\Psi^-} = - \ket{\Psi^-}$,
$\sigma_{x}^A \sigma_{x}^B \ket{\Psi^-} = - \ket{\Psi^-}$.
This 
suggests that the way to pass from discrete variables to continuous
variables is to replace products of Pauli matrices by sums of position 
and momentum operators. But then one does not see how to obtain a GHZ
paradox of the form eq. (\ref{spinGHZ}) 
since addition of operators is always commutative. We shall
show that their are
two equivalent ways to circumvent this. The first is to work
with products of 
translation operators of the form $\exp(i \alpha x)$ and $\exp(i
\beta p)$; the second is to work with sums of
modular variables (first introduced 
in \cite{a}). 
The origin of the paradox is in one case 
the 
non-commutativity of translation operators
and in the second the non associativity of the modulo operation
for operators. Finally we introduce a new kind of variable, which we
call binary variables, in terms of which the continuous variable GHZ
paradox can be mapped onto the 
the GHZ paradox for spins eq. (\ref{spinGHZ}).

Multi-particle non local states for continuous variables
have been considered previously by van Loock and
Braunstein\cite{vLB}. One of the interests of the states considered by 
van Loock and Braunstein is that they
can be
easily constructed using squeezed states and beam splitters. But the
measurements that exhibit the non locality are complicated and cannot
be realized in the laboratory at present. On the
other hand the states we discuss here seem significantly more complicated to
construct than those considered by van Loock and
Braunstein (in particular they cannot be constructed using squeezers
and beam splitters). But the measurements that exhibit the non locality are
simple position and momentum measurements.

Let us consider a set of unitary operators $X_j$, $Y_j$, $j=A,B,C$
acting on A,B and C's particles respectively. Using these operators we 
construct the following 4 GHZ operators:
\begin{eqnarray}
V_1 &=& X_A X_B X_C\ , \nonumber\\
V_2 &=& X_A^\dagger Y_B Y_C^\dagger \ ,\nonumber\\
V_3 &=& Y_A^\dagger X_B^\dagger Y_C \ ,\nonumber\\
V_4 &=& Y_A Y_B^\dagger X_C^\dagger \ .
\label{V4}
\end{eqnarray}
These four operators give rise to a GHZ paradox if 
$X_j$ and $Y_j (Y_j^\dagger)$ anticommute:
\begin{equation}
X_j Y_j + Y_j X_j = 0 \quad {\mbox{and}} \quad
X_j Y_j^\dagger + Y_j^\dagger X_j = 0\ .
\label{aa}
\end{equation}
This implies that the following two properties hold:
\begin{enumerate}
\item $V_1, V_2, V_3, V_4$ all commute. This ensures that they can be
  simultaneously diagonalized (in fact there exists a complete set
  of common eigenvectors).
\item The product $V_1 V_2 V_3 V_4 =
-I_{ABC} $ equals minus the   identity operator.
 \end{enumerate}
Any common eigenstate of $V_1, V_2, V_3, V_4$ will then give rise to
a GHZ paradox. 
Indeed suppose that parties measure the hermitian operators $-i\ln X_j$ or
$-i\ln Y_j$, $j=A,B,C$ on this common eigenstate. The result of the
measurement associates a complex number of unit norm to either the
$X_j$ or $Y_j$ operators. Hence if one of the  combinations of operators
that occurs in eq. (\ref{V4}) is measured, 
one can assign a value to one of the operators $V_1, V_2,
V_3, V_4$. Quantum mechanics imposes that this value is equal to the
corresponding eigenvalue. 

But in a local hidden variables theory this is
impossible because in this case one must assign, prior to the
measurement, a complex number of unit norm to all the operators $X_j$ and
$Y_j$. But then taking the product of the four c-numbers assigned
simultaneously to 
$V_1, V_2,
V_3, V_4$ yields $+1$ whereas -due to property 2- the product of the
eigenvalues is $-1$. 

In order to construct such unitary operators in the continuous case,
let us introduce the dimensionless variables 
\begin{equation}
\tilde x = {x \over {\sqrt{\pi}L}} \quad , \quad
\tilde p = {p\ L \over \sqrt{\pi}}
\end{equation}
where $L$ is an arbitrary length scale.
Upon taking 
\begin{equation}
X_j = \exp ( i \pi \tilde x_j) \quad \mbox{and} \quad
Y_j = \exp ( i \pi \tilde p_j)\ ,\  j=A,B,C
\label{XY}
\end{equation}
 one finds that
the anticommutation relations eq. (\ref{aa})  are
satisfied. This follows from the
identity 
\begin{eqnarray}
&&\exp ( i \pi \tilde x_j \alpha) \exp ( i \pi \tilde p_j \beta)\nonumber\\
&& =  \exp ( i \pi \tilde p_j \beta) \exp ( i \pi \tilde x_j \alpha)
\exp(- i \pi \alpha \beta) \ .
\label{expcomm}
\end{eqnarray}
In particular we see that if 
 the product $\alpha \beta$ is an even integer, 
$\exp ( i \pi \tilde x_j \alpha)$ and $ \exp ( i \pi \tilde p_j \beta)$
commute, whereas if $\alpha \beta$ is an odd integer they anti-commute.

Thus any common eigenstate of $V_1, V_2, V_3, V_4$ 
with $X_j, Y_j$ given in eq. (\ref{XY})
yields a GHZ
paradox if the three parties carry out local measurements of either
$x$ or $p$. Before describing the states that yield the GHZ paradox,
we pursue the algebraic analysis of these equations.

By taking the logarithm of the GHZ equations and dividing by $i\pi$, we obtain
the following form in terms of hermitian operators
\begin{eqnarray}
( \tilde x_A + \tilde x_B + \tilde x_C ) \mbox{mod} 2
\ket{\Psi_{GHZ}'} &=& \eta_1\ket{\Psi_{GHZ}'} ,
\nonumber\\
( - \tilde x_A + \tilde p_B - \tilde p_C ) \mbox{mod} 2
\ket{\Psi_{GHZ}'} &=& \eta_2 \ket{\Psi_{GHZ}'} ,
\nonumber\\
( - \tilde p_A - \tilde x_B + \tilde p_C ) \mbox{mod} 2
\ket{\Psi_{GHZ}'} &=& \eta_3 \ket{\Psi_{GHZ}'} ,
\nonumber\\
( \tilde p_A - \tilde p_B - \tilde x_C ) \mbox{mod} 2
\ket{\Psi_{GHZ}'} &=& \eta_4 \ket{\Psi_{GHZ}'} ,
\label{modularGHZ}
\end{eqnarray}
where the eigenvalues $\eta_k \in [0, 2[$ obey the
relation
\begin{equation}
( \eta_1 + \eta_2 + \eta_3 + \eta_4 ) \mbox{mod} 2 = 1
\label{eta}
\end{equation}
We recall that if $ \ket{a}$ is an eigenvector of the operator $A$
with eigenvalue $a$, then the modular operator
$(A) \mbox{mod} k$ is defined by 
$(A) \mbox{mod} k \ket{a} = (a) \mbox{mod} k\ket{a} $.  
Modular variables where introduced in \cite{a} as a general
tool to study non-locality in quantum mechanics. It is interesting that
they reappear in the context of the GHZ paradox.

In the case of eq. (\ref{modularGHZ}), 
the paradox arises because the associativity of the modulo operation
$(A + B) \mbox{mod} k = 
((A)\mbox{mod} k + (B) \mbox{mod} k ) \mbox{mod} k$
which holds for c-numbers is in general not valid when $A$ and $B$ are
 non-commuting
operators. 
Thus
in a local hidden variable theory one must assign real values to $(\tilde
x_j)\mbox{mod} 2 $ and to $(\tilde p_j)\mbox{mod} 2 $. Then taking the sum of
the 4 equations (\ref{modularGHZ}) and using the associativity property of
modulo for c-numbers one finds $( \eta_1 + \eta_2 + \eta_3 + \eta_4 )
\mbox{mod} 2 = 0$ in contradiction with the quantum condition eq. (\ref{eta}).

We shall now rephrase the GHZ paradox
eq. (\ref{modularGHZ}) in a different way by using binary variables.
Consider a position eigenstate $\ket{x}$ and write its
eigenvalue in base 2 as
\begin{equation}
x = (-1)^{\mbox{sgn$x$}} {L \sqrt{\pi}} \sum_{n=
  -\infty}^{+\infty} [\tilde x]_n 2^{n} \ .
\label{xbin}
\end{equation}
This allows us to introduce the sign operator $\widehat {{\mbox{sgn}x}}$ and
the binary operators $\widehat {[\tilde x]_n}$ defined by
\begin{eqnarray}
\widehat {{\mbox{sgn}x}} \ket{x} &=& \mbox{sgn}x \ket{x} , \nonumber\\
\widehat {[\tilde x]_n}  \ket{x} &=& [\tilde x]_n \ket{x}\ .
\end{eqnarray}
The modular position is then written as
\begin{equation}
(\tilde x) {\mbox{mod}2^k} = \sum_{n=-\infty}^{k-1} 
\widehat { [\tilde x]_n} 2^{n}
\label{modpos}
\end{equation}
Similarly we can introduce the operators  $\widehat {{\mbox{sgn}p}}$ and
 $\widehat {[\tilde p]_n}$ using the 
base 2 decomposition of momentum
\begin{equation}
p = (-1)^{\mbox{sgn$p$}} {\sqrt{\pi} \over L} \sum_{n=
  -\infty}^{+\infty} [\tilde p]_n 2^{n} \ .
\label{pbin}       
\end{equation}
Using these definitions we have the relation 
\begin{equation}
(z){\mbox{mod}{2^{k+1}}} = (z){\mbox{mod}{2^{k}}} + [z]_k 2^k
\label{z}
\end{equation}
for
$z= \tilde x, \tilde p$ (from now on we drop the $\hat{ }$ over operators 
since it will be clear from the context whether $x, p$ denote
operators or c-numbers).
We can then rewrite eq. (\ref{modularGHZ}) 
as
\begin{eqnarray} 
&&\left( (\tilde x_A)\mbox{mod}1 + (\tilde x_B)\mbox{mod}1 + (\tilde
x_C)\mbox{mod}1 \right.\nonumber \\
& & \quad \left.+ [ \tilde x_A]_0 + [\tilde x_B]_0 + [\tilde x_C]_0
\right)\mbox{mod}2 \ \ket{\Psi_{GHZ}'}
%\nonumber\\
%&=&  
=\eta_1\ket{\Psi_{GHZ}'} \ ,\nonumber\\
&&\left( -(\tilde x_A)\mbox{mod}1 + (\tilde p_B)\mbox{mod}1 - (\tilde
p_C)\mbox{mod}1 \right. \nonumber \\
& &\quad \left.  + [ \tilde x_A]_0 + [\tilde p_B]_0 + [\tilde p_C]_0
\right)\mbox{mod}2 \ \ket{\Psi_{GHZ}'}
%\nonumber\\
%&=&  
=\eta_2\ket{\Psi_{GHZ}'} \ , \nonumber\\
&&\left( -(\tilde p_A)\mbox{mod}1 - (\tilde x_B)\mbox{mod}1 + (\tilde
p_C)\mbox{mod}1 \right.\nonumber \\
& & \quad \left.+ [ \tilde p_A]_0 + [\tilde x_B]_0 + [\tilde p_C]_0
\right)\mbox{mod}2 \ \ket{\Psi_{GHZ}'}
%\nonumber\\
%&=&  
=\eta_3\ket{\Psi_{GHZ}'}  \ ,\nonumber\\
&&\left( (\tilde p_A)\mbox{mod}1 - (\tilde p_B)\mbox{mod}1 - (\tilde
x_C)\mbox{mod}1\right. \nonumber \\
& & \quad \left.+ [ \tilde p_A]_0 + [\tilde p_B]_0 + [\tilde x_C]_0
\right)\mbox{mod}2 \ \ket{\Psi_{GHZ}'}
%\nonumber\\
%&=&  
=\eta_4\ket{\Psi_{GHZ}'} \ , \nonumber\\
\label{GHZmodbin}
\end{eqnarray}
where we have used the fact that $[-z]_0=[z]_0$.

In order to understand the structure of eq. (\ref{GHZmodbin}),
we note that (\ref{expcomm}) implies that
$(\tilde x){\mbox{mod}{2^k}}$ and
$(\tilde p){\mbox{mod}{2^l}}$ commute if $k + l \leq 1$. Indeed, $(\tilde
x){\mbox{mod}{2^k}}$ and $e^{i{2 \over 2^k}\pi \tilde x }$ have the same
eigenstates and similarly $(\tilde p){\mbox{mod}{2^l}}$ and $e^{i {2 \over
2^l}\pi \tilde p}$. Thus if 
$e^{i{2 \over 2^k}\pi \tilde x }$ and
$e^{i {2 \over 2^l}\pi \tilde p}$ commute, 
they share a common basis of
eigenstates, and therefore 
$(\tilde x){\mbox{mod}{2^k}}$ and
$(\tilde p){\mbox{mod}{2^l}}$ 
do too. But from (\ref{expcomm}), 
 $e^{i{2 \over 2^k}\pi \tilde x }$ and $e^{i {2 \over 2^l}\pi \tilde p}$
commute only if $k+l\leq 1$.
Using (\ref{z}), we also have that $(\tilde x){\mbox{mod}{2^k}}$ and ${[\tilde
p]_m}$ commute if $k + m \leq 0$; $(\tilde p){\mbox{mod}{2^l}}$ and ${[\tilde
x]_m}$ commute if $l + m \leq 0$; ${[\tilde x]_n}$ and ${[\tilde
p]_m}$ commute if $n + m \leq -1$.

From these properties we deduce that the mod1
terms 
on the left hand side
of the four equations (\ref{GHZmodbin}) commute with all the
other terms that appear in these equations. These terms are therefore not
essential to the paradox and can be dropped. Ommiting them yields the
following simple form for eq. (\ref{GHZmodbin})
\begin{eqnarray}
 ([ \tilde x_A]_0
+  [\tilde x_B]_0 + [\tilde x_C]_{0})\mbox{mod}2  \ket{\Psi_{GHZ}} &=&
b_1\ket{\Psi_{GHZ}}, \nonumber\\
([ \tilde x_A]_0
+  [\tilde p_B]_0 + [\tilde p_C]_{0})\mbox{mod}2  \ket{\Psi_{GHZ}} &=&
b_2\ket{\Psi_{GHZ}}, \nonumber\\
 ([ \tilde p_A]_0
+  [\tilde x_B]_0 + [\tilde p_C]_{0})\mbox{mod}2  \ket{\Psi_{GHZ}} &=&
b_3\ket{\Psi_{GHZ}}, \nonumber\\
([ \tilde p_A]_0
+  [\tilde p_B]_0 + [\tilde x_C]_{0})\mbox{mod}2  \ket{\Psi_{GHZ}} &=&
b_4\ket{\Psi_{GHZ}}, \nonumber\\
\label{binary1GHZ}
\end{eqnarray}
where $b_{1,2,3,4} \in \{0,1\}$ satisfy
\begin{equation}
(b_1+b_2+b_3+b_4)\mbox{mod}2 =1
\end{equation}

Once again the paradox is due to the associativity of
the  modulo operation which holds for c-numbers but does not
hold for operators. (As an example the sum of two even integers is an
even integer, but if one takes the sum of two operators 
both of whose eigenvalues 
are even integers, one does not in general obtain an operator whose
eigenvalues are only even integers).

Exponentiating eq. (\ref{binary1GHZ}) we obtain
\begin{eqnarray}
e^{i\pi{[ \tilde x_A]_{0}}}
 e^{i\pi{[ \tilde x_B]_{0}}}
e^{i\pi {[ \tilde x_C]_{0}}}
\ket{\Psi_{GHZ}} &=& e^{i\pi b_1}\ket{\Psi_{GHZ}},
\nonumber\\
 e^{i\pi {[ \tilde x_A]_{0} }}
e^{i\pi {[ \tilde p_B]_{0} }}
e^{i\pi {[ \tilde p_C]_{0} }}
\ket{\Psi_{GHZ}} &=& e^{i\pi b_2}\ket{\Psi_{GHZ}},
\nonumber\\
 e^{i\pi {[ \tilde p_A]_{0} }}
e^{i\pi {[ \tilde x_B]_{0} }}
e^{i\pi {[ \tilde p_C]_{0} }}
\ket{\Psi_{GHZ}} &=& e^{i\pi b_3}\ket{\Psi_{GHZ}},
\nonumber\\
e^{i\pi  {[ \tilde p_A]_{0}  }}
e^{i\pi {[ \tilde p_B]_{0} }}
e^{i\pi {[ \tilde x_C]_{0}  }}
\ket{\Psi_{GHZ}} &=& e^{i\pi b_4}\ket{\Psi_{GHZ}} .
\label{binaryGHZ}
\end{eqnarray}
We now show that 
this is identical to the original GHZ paradox for spins. Indeed the
operators $ X= e^{i\pi \widehat {[ \tilde x]_{0} }}$, $Y= e^{i\pi \widehat {[
\tilde p]_{0} }}$, and $Z = -i XY$ are a representation of $su(2)$ that obey
the usual commutation relations $[X,Y]=2 i Z$ and cyclic permutations. This can
be verified using the eigenstates of $Z$:
\begin{eqnarray}
\ket{\uparrow}_{\tilde x_0, \tilde p_0}
&=& {1 \over \sqrt{2}}
\left( 
\sum_{k=-\infty}^{\infty} e^{i\pi 2k\tilde p_0} \ket{\tilde x =
\tilde x_0 +2k}\right. \nonumber\\
&&
\left.
+i\sum_{k=-\infty}^{\infty}e^{i\pi (2k+1)\tilde
p_0}\ket{\tilde x= \tilde x_0+2k+1} \right) 
\nonumber\\
&=& {e^{-i \tilde x_0 \tilde p_0} \over \sqrt{2}}
\left( 
\sum_{k=-\infty}^{\infty}
e^{-ik\pi\tilde x_0} \ket{\tilde p = \tilde p_0 +k}
\right.
\nonumber\\
& &
\left.
+i\sum_{k=-\infty}^{\infty}e^{-ik\pi(\tilde x_0+1)}\ket{\tilde p= \tilde
p_0+k} 
\right) 
\end{eqnarray}
and
\begin{eqnarray}
\ket{\downarrow}_{\tilde x_0, \tilde p_0} &=& 
{1 \over \sqrt{2}}\left(
\sum_{k=-\infty}^{\infty} {1 \over \sqrt{2}}
e^{i\pi 2k\tilde p_0} \ket{\tilde x = \tilde x_0 +2k} \right.
\nonumber\\
&& \left.
-i\sum_{k=-\infty}^{\infty}e^{i\pi (2k+1)\tilde p_0}\ket{\tilde x= \tilde
x_0+2k+1} \right)  \nonumber\\ 
&= & {e^{-i \tilde x_0 \tilde p_0} \over \sqrt{2}}
\left( 
\sum_{k=-\infty}^{\infty}
e^{-ik\pi\tilde x_0} \ket{\tilde p = \tilde p_0 +k}
\right.\nonumber\\
&&\left.
-i\sum_{k=-\infty}^{\infty}e^{-ik\pi(\tilde x_0+1)}\ket{\tilde p= \tilde
p_0+k}\right)
\end{eqnarray}
where $\tilde x_0$ and $\tilde p_0 \in
[0,1[$ and where we have defined \begin{equation} \ket{\tilde
x}=\ket{x=\sqrt{\pi}L\tilde x}\quad , \quad
\ket{\tilde
p}=\ket{p=\sqrt{\pi}\tilde p/L}
 \end{equation}
These states form a basis of the Hilbert space. The action of $X$, $Y$ and $Z$
on them is \begin{eqnarray}
X \ket{\uparrow}_{\tilde x_0, \tilde p_0} = \ket{\downarrow}_{\tilde x_0,
\tilde p_0}
&,&
X \ket{\downarrow}_{\tilde x_0, \tilde p_0} =  \ket{\uparrow}_{\tilde x_0,
\tilde p_0},
\nonumber\\
Y \ket{\uparrow}_{\tilde x_0, \tilde p_0} =i \ket{\downarrow}_{\tilde x_0,
\tilde p_0},
&,&
Y \ket{\downarrow}_{\tilde x_0, \tilde p_0} = -i \ket{\uparrow}_{\tilde x_0,
\tilde p_0},
\nonumber\\
Z \ket{\uparrow}_{\tilde x_0, \tilde p_0} = \ket{\uparrow}_{\tilde x_0,
\tilde p_0},
&,&
Z \ket{\downarrow}_{\tilde x_0, \tilde p_0} = - \ket{\downarrow}_{\tilde x_0,
\tilde p_0} .
\nonumber\\
\end{eqnarray}
This shows that the GHZ paradox we have constructed
for continuous variables is not fundamentally different from the original
paradox. It is only expressed in terms of an $su(2)$ sub-algebra of an infinite
dimensional Hilbert space.

To conclude we now describe the common eigenstate of the
four GHZ operators in the case of eqs.(\ref{modularGHZ}) and (\ref{binaryGHZ}).
Define
\begin{eqnarray}
\ket{b_i}_j&=\ket{\uparrow}_{\tilde x^j_0, \tilde
p^j_0}&\mbox{if}\ b_i\mbox{mod}2=0
\nonumber\\
&=\ket{\downarrow}_{\tilde x^j_0, \tilde
p^j_0}&\mbox{if}\ b_i\mbox{mod}2=1
\end{eqnarray}
where $j=A,B,C$. Using this notation, the state
\begin{eqnarray}
\ket{\psi (\bf{b},\bf{z})}&=& {1 \over \sqrt{2}} \left (
\ket{b_2}_A\ket{b_3}_B\ket{b_4}_C
\right.
\nonumber\\
& & 
\left.
+ (-1)^{b_1}\ket{b_2+1}_A\ket{b_3
+ 1 } _B \ket{ b _ 4+1}_C \right)
\label{B}
\end{eqnarray}
depending on the variables 
${\bf b}= (b_1, b_2,b_3,b_4)$ and
${\bf z}=(\tilde x^A_0,\tilde x^B_0,\tilde
x^C_0,\tilde p^A_0,\tilde p^B_0,\tilde p^C_0)$ is a solution of
eq. (\ref{binaryGHZ}). The general eigenstate of eq. (\ref{binaryGHZ})
is then of the form
\begin{equation} \ket{\Psi_{GHZ}({\bf b})}=
\int d{\bf z}\  f({\bf z})\ket{\psi ({\bf b},{\bf z})}
\end{equation} where $f({\bf z})$ is some normalized function.

Finally, we can use this expression to obtain the eigenstates of
 eq (\ref{modularGHZ}) (or equivalently
(\ref{GHZmodbin})). It is easy to check that
\begin{eqnarray}
&&\left[
(z_A)\mbox{mod}1 + (z_B)\mbox{mod}1 + (z_C)\mbox{mod}1 \right]
\ket{\psi (\bf b,z)}
\nonumber\\
&=& \left[ z_0^A + z_0^B + z_0^C \right]
\ket{\psi (\bf b,z)}
\end{eqnarray}
where $z$ stands for $\tilde x$ or $\tilde p$. 
Therefore inserting the state (\ref{B}) into eq. (\ref{GHZmodbin}) we obtain
the equations
\begin{eqnarray}
(\tilde x_0^A+\tilde x_0^B+\tilde x_0^C+b_1)\mbox{mod}2&=&\eta_1
\nonumber \\
(-\tilde x_0^A+\tilde p_0^B-\tilde
p_0^C+b_2)\mbox{mod}2&=&\eta_2 \nonumber \\
(-\tilde
p_0^A-\tilde
x_0^B+\tilde p_0^C+b_3)\mbox{mod}2&=&\eta_3
\nonumber \\
(\tilde
p_0^A-\tilde
p_0^B-\tilde x_0^C+b_4)\mbox{mod}2&=&\eta_4 \ .
\label{contr}
\end{eqnarray}
 The general solution of 
eq. (\ref{GHZmodbin}) is then an arbitrary superposition of states 
(\ref{B}) for which $
\bf b,z$ obey the constraints (\ref{contr}).
If we denote by $\Xi$ the set of solutions of
(\ref{contr}), then the general solution can be written
\begin{equation}
\ket{\Psi_{GHZ}'}= \int_\Xi d\mbox{\boldmath{$\xi$}}\
g(\mbox{\boldmath{$\xi$}}) \ \ket{\psi (\mbox{\boldmath{$\xi$}})}
\end{equation} where $\mbox{\boldmath{$\xi$}}=\bf ( b , z )$.

In summary we have exhibited a large class  of states that satisfy a GHZ
paradox. Exhibiting the paradox is very simple: it requires only that
the parties carry out local measurements of position and
momentum. However constructing these states is a much more difficult
task. For instance it seems impossible to even approximately  construct
them  using squeezers and beam splitters, contrary to the states
described in \cite{vLB}. Constructing them explicitly will therefore
require more sophisticated quantum state engineering than we
presently possess.

{\bf Acknowledgements:} S.M. is a research associate of the Belgian
National Research Foundation. He acknowledges funding by the European
Union under project EQUIP (IST-FET program).

\end{multicols}

\end{document}